\documentclass[twocolumn]{aastex631}
\usepackage{amsmath}
\usepackage{subfigure}
\usepackage{xcolor}
\usepackage{float,color}
\usepackage{gensymb}
\usepackage{placeins}
\usepackage{multirow}
\usepackage{appendix}

\usepackage{mdframed}

\begin{document}

\title{Early warning for lensed gravitational wave counterparts from time delays of their host galaxies observed in the optical}
\shorttitle{Early warning of lensed GW counterparts}


\author[0000-0002-0870-2993]{Sourabh Magare}
\affiliation{Inter-University Centre for Astronomy and Astrophysics, Post Bag 4, Ganeshkhind, Pune - 411007, India}
\correspondingauthor{Sourabh Magare}
\email{sourabh.magare@gmail.com}

\author[0000-0001-7714-7076]{Anupreeta More}
\affiliation{Inter-University Centre for Astronomy and Astrophysics, Post Bag 4, Ganeshkhind, Pune - 411007, India}
\affiliation{Kavli IPMU (WPI), UTIAS, The University of Tokyo, Kashiwa, Chiba 277-8583, Japan}

\author[0000-0001-5318-1253]{Shasvath J. Kapadia}
\affiliation{Inter-University Centre for Astronomy and Astrophysics, Post Bag 4, Ganeshkhind, Pune - 411007, India}
\shortauthors{Magare, More \& Kapadia}

\begin{abstract}
Gravitational lensing of gravitational waves (GWs) can be leveraged to provide early-warning times of $\mathcal{O}({\rm hours})$ to $\mathcal{O}({\rm days})$ before the merger of Binary Neutron Stars (BNSs) and Neutron Star Black Holes (NSBHs). This in turn could enable electromagnetic (EM) telescopes to capture emissions surrounding the time of the merger.
In this work, we assess the practicability of lensing-driven early-warning by analysing optical images of the lensed host galaxy to predict the arrival time of subsequent BNS/NSBH signals following the observation of the first signal. We produce mock lenses with image quality and resolution similar to images taken with the Hubble Space Telescope (HST) and the ground-based Hyper Suprime-Cam (HSC) on the Subaru telescope. We compare the time delay uncertainties between these two cases for typical lensed image configurations and multiplicity. These include doubles and quads, and among quads: the fold, cusp, cross image configurations. We find that time delay uncertainties for doubles are comparable for both HST and HSC mocks. On the other hand, quads tend to provide accurate time-delay predictions (typical relative error $\sim0.1$) with HST. Analysis of a real lens led to a difference in time-delay estimates of $\mathcal{O}(\rm days)$ between the predictions derived from HST and HSC data. Our work therefore strongly advocates the need for high-resolution EM observations of lensed host galaxies to feasibly enable lensing-driven early-warning.
\end{abstract}

\section{Introduction} \label{sec:intro}
The advent of multi-messenger astronomy using gravitational waves (GWs) began with the discovery of GW170817 by the LIGO-Virgo collaboration \citep{PhysRevLett.119.161101, Abbott_2017_apjL}. This was the first detection of a binary neutron star (BNS) merger that also had an electromagnetic (EM) counterpart and was extensively followed by telescopes around the world and across the entire EM spectrum. This detection enabled unique constraints in fundamental physics, e.g. upper limits on the mass of the graviton \citep{PhysRevD.100.104036}, bounds on the speed of GWs \citep{Abbott_2017_apjL}, constraints in Lorentz invariance \citep[e.g.][]{PhysRevD.97.124023, PhysRevD.97.084040}, probing the violation of the equivalence principle \citep{Wei_2017}; shed light on the provenance of hitherto unknown short GRBs (sGRBs) and Kilonovae (KNe) \citep{Abbott_2017_apjL}; and showed BNSs to be sites where certain heavier elements of the periodic table are synthesized \citep{Kasen:2017sxr}. 

Although, GW170817 was a seminal event, there was nevertheless a delay between the detection of GWs and follow-up by the EM telescopes \citep{Abbott_2017_apjL}. This motivates the need for an early warning of the merger to capture the pre-merger \citep[e.g][]{PhysRevLett.108.011102, Most_2020}, merger, and post-merger EM emissions \citep{Siegel_2016}. Early optical and UV spectra can allow us to map the mass and the velocity function of the ejecta and refine our understanding of r-process nucleosynthesis \citep{Nicholl_2017}. An early x-ray observation can inform about the post-merger state of the remnant \citep{Siegel_2016}. An early radio spectrum observation could test the theories that identify the magnetic interaction with BNS merger as the source of fast radio bursts \citep{Most2020ElectromagneticPT, Wang_2016}.  A pre-merger observation of EM emissions promises to answer longstanding questions in neutron star astrophysics. 

A crucial requirement for observing premerger EM emissions is to localize the source before their onset. Current methods for GW early warning \citep{Cannon_2012} rely on accumulating a sufficient signal-to-noise ratio (S/N) before the merger, and constructing a sky-localization bound only based on this inspiral part of GWs. Further, since BNS events spend a relatively longer time (as compared to binary black hole systems) in the LIGO-Virgo-Kagra (LVK) \citep{Aasi_2015, Acernese_2015, 2018LRR....21....3A, 10.1093/ptep/ptaa125, PhysRevD.102.062003} frequency band, this technique seems to be better suited for BNS mergers. The typical early warning time for LVK in O4 observing run is $\sim10$ seconds for sky localization of $\mathcal{O}(100)$ sq deg \citep{Sachdev_2020, Magee_2021}. Although the early warning time increases in O5 and Voyager observing run to $\sim30$ sec and $\sim1$ min respectively \citep{2018LRR....21....3A, adhikari2018ligo}, it is likely insufficient to identify and detect the EM emissions arising from the pre-merger and merger of the BNS events. For neutron star black hole (NSBH) mergers, early-warning times are reduced because of their relatively higher masses reducing the in-band duration. Including higher harmonics of GWs in real time searches \citep{Kapadia_2020, 10.1093/mnras/stab125, 10.1093/mnras/stac852}, can purchase an additional $\mathcal{O}(10)$ sec in second generation GW detectors, but will likely remain insufficient to capture EM counterparts surrounding the merger.

In our earlier work \citep{Magare_2023}, we proposed a novel astrophysical scenario by which we can get an early warning of $\mathcal{O}({\rm hours})$ to $\mathcal{O}({\rm days})$ before the merger, that too with a sub-arc second sky-localization, provided the BNS/NSBH is strongly lensed. The scenario is that if we detect GWs from BNS or NSBH event that have an EM counterpart, then just like GW170817, by the EM follow-up searches we can localize the merger to its host galaxy. If we find that the host galaxy is strongly lensed, then using the position of the images of the host galaxy, along with the information of the redshift of the host and the lens galaxy, we can predict the arrival time of the next image. This arrival time of the images will act as the early warning of the BNS/NSBH merger and thus the early warning time will be drastically enhanced. 

From this scenario, we estimated that up to $\sim 5 \%$ of detectable lensed BNSs in O4 will have early warning time of $1$ day. This percentage increases to $\sim 10 \%, \sim 45 \%$ and $\sim 90 \%$ in O5, Voyager and 3G observing runs respectively \citep[see][]{Magare_2023}. For NSBHs, we get more optimistic results. We find that $\sim 10 \%$ of detectable lensed NSBHs in O4 will have an early warning time of $1$ day. For future observing runs this percentage increases to $\sim 35 \%, \sim 70 \%$ and $\sim 95 \%$ in O5, Voyager and 3G respectively.
Further, our prediction of the time delays has errors ranging from a few hours to a few days depending on the position of the source in the lens plane, as well as the redshifts of the source and lens galaxies. This was based on assuming that the images have an angular resolution of $0.05''$ arc seconds in optical wavelengths, which is the resolution of a space telescope or a ground based telescope using adaptive optics.  

However, in our earlier work we made some simplifying assumptions. These need to be addressed to assess the practicability of lensing driven early warning. We assumed images of the background/host galaxy as point sources and the lensed images were assumed to have astrometric uncertainties similar to that from Hubble Space Telescope (HST, $1\sigma=0.05''$). However, most lensed galaxy images are extended and it may not always be possible to use a space telescope to follow up every EM counterpart. 
It is, therefore, of interest to check if lensing driven early warning can be realized with a ground-based telescope as well. 

In this work, we attempt to forgo these assumptions to make our analysis more realistic. We simulate realistic images of lens systems and assume S\'ersic light profiles for both the foreground and background galaxies. For each mock lens, we produce images to mimic the image quality and resolution of HST (space-based imaging) \citep{2012OptEn..51a1011L,2023acsi.book...23R}  and that of Hyper Suprime-Cam (HSC) on Subaru Telescope (ground-based imaging) \citep{10.1093/pasj/psx063,2018PASJ...70S...4A}.  Given these mock samples, we perform the lens light subtraction and then fit the lens model to infer the parameters of the lens system and hence the time delays between the images. We use the software \texttt{glafic} \citep{glafic_ref} for creating the images of simulated lens systems. We then do the lens light subtraction to make the images more contrasting and easily identifiable. We use the package \texttt{imcascade} \citep{Miller_2021} for lens light subtraction. Using this residual image, we use Singular Isothermal Ellipsoid (SIE) with external shear \citep{1991ApJ_373_354K, 1994A&A_284_285K} as the lens model and infer the parameters of the lens and host galaxy using the Differential Evolution (DE) algorithm \citep{journals_jgo_StornP97,Qiang2014AUD}. Once we find the best-fit parameters, we get time delay predictions for both HST and HSC mock lenses. We compare the estimated parameters and uncertainties in the time delays because of the different telescope resolutions. We also apply this method to a real lens system which is observed in both of these telescopes.

The paper is organized as follows: Section \ref{methods_section} gives a description of preparation of simulated images, lens light subtraction, and parameter estimation using the differential evolution algorithm. Section \ref{results} describes the results and compares the time delays predictions for mock HST and HSC data. Section \ref{conclusion_section} gives the summary and the conclusion of the work.

\begin{figure*}
    \includegraphics[scale = 0.45 ]{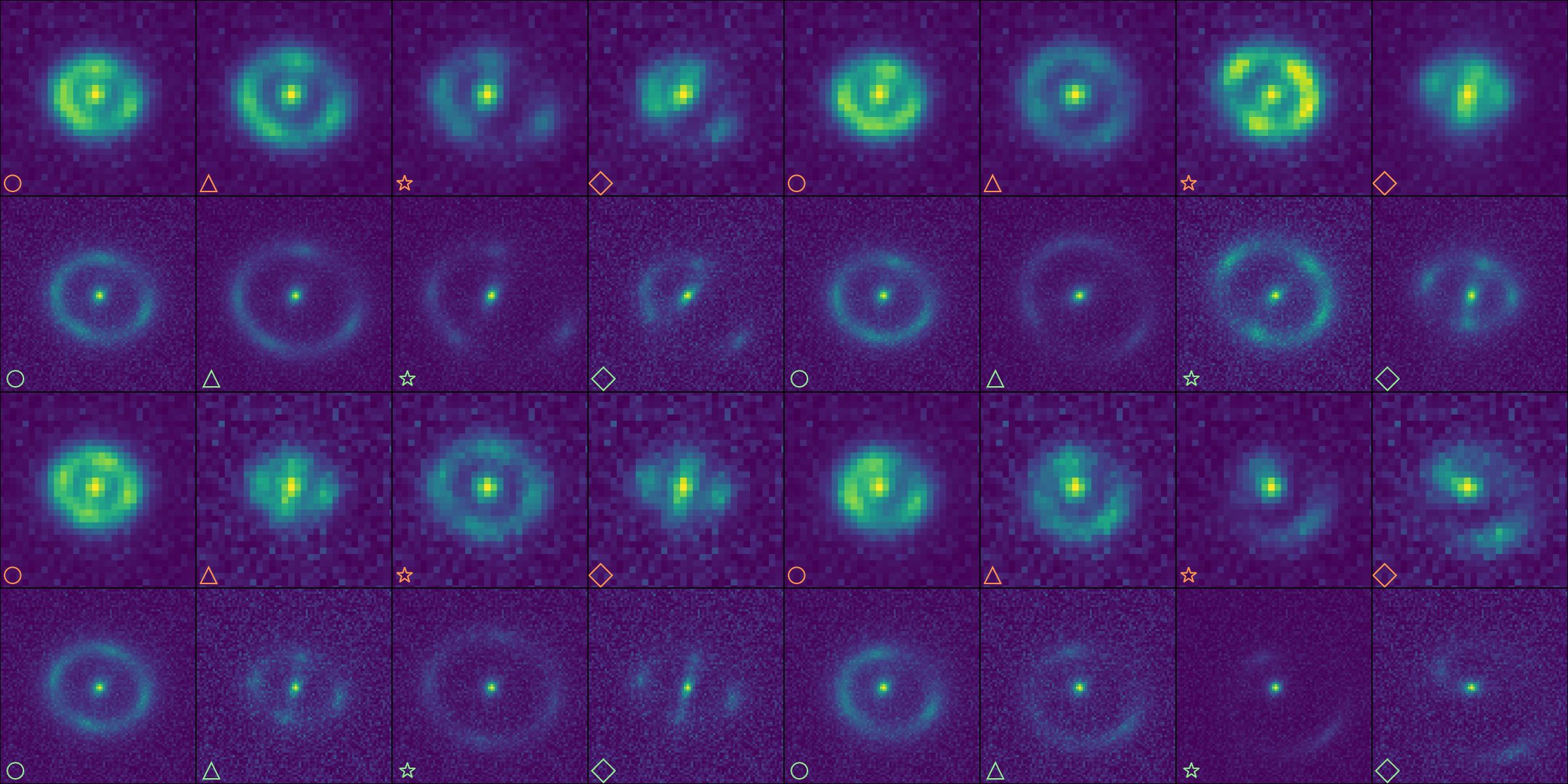}
    \caption{All 16 simulated lens systems shown for the mock HST (\textit{green}) and the corresponding HSC (\textit{orange}) imaging. Each lens system is shown with a marker and corresponds to the same lens shown in other figures.
    }
    \label{fig:Lens_images}
\end{figure*}

\section{Methods}\label{methods_section}
In our earlier work \citep{Magare_2023}, we assumed point-like lensed images and that the measurement uncertainties follow a Gaussian distribution with a standard deviation of $0.05''$. This is comparable to pixel resolutions achieved by space-based telescopes or ground-based telescopes with adaptive optics. However, many of the lensed sources are extended galaxies producing extended arc-like images with added complexities, for instance, multiple arcs that are overlapping \citep[e.g.][]{1988ApJ_324L_37G, 1996ApJ_473L_17D}. This can lead to incorrect identification of the various counterparts of lensed images or inaccuracies in measuring their positions. Another complication is the observability of the lensed images. Often the lens galaxy is much brighter than the lensed images or the lensed images closer to the center of the lens are demagnified making them inherently fainter. This further contributes to uncertainties in measuring their positions accurately. Here, we study how the aforementioned factors will impact the time delay predictions by simulating both high and low resolution lens systems, performing lens-light subtraction and modeling the system.

\subsection{Generating Simulations}\label{ssec_gen_sim}
We use \texttt{glafic} to simulate images of lens systems \citep{glafic_ref}. 
For generating the lensed images, we adopt a mass model given by SIE and an external shear and assume light distribution follows mass distribution. The lens light is assumed to follow de Vaucouleur's profile (i.e.  S\'ersic index ($n$) is set to 4), as elliptical galaxies are well fitted by this profile. The center of the lens mass (and light) is located at the center of the image. 
The source galaxy is also modeled by the S\'ersic profile with $n$ of $1.5$, which corresponds to typical spiral and lenticular galaxies. In \citep{Magare_2023}, we had drawn lensed source redshifts, intrinsic source parameters e.g. masses of the BNS/NSBH, lens redshifts, intrinsic lens parameters e.g. velocity dispersion ($\sigma_{\rm l}$) and also extrinsic source parameters from realistic populations. We checked the detectability of GWs from such systems for O$5$ observing run \citep{Magare_2023}, with the detectability criterion that any two lensed images have SNR greater than 8. We draw parameters of our lens systems from these existing lens samples. 

In our simulations, we generate doubly lensed sources (doubles) and the three standard image configurations of quadruply lensed sources (quads), namely, a fold, a cusp and a cross. These typical lensed image configurations are qualitatively distinct and are expected to show systematic differences in the accuracy of the time delay predictions.  
For each configuration, we simulate four randomly generated lenses and each lens system is created at high and low angular resolutions corresponding to the HST and HSC imaging, respectively.  
Thus, we have a  total of $16$ unique lens systems combining the quads and doubles. 
The mock HST and HSC images have pixel resolutions of $0.0445$ and $ 0.168$ arcsecond per pixel. We generate Gaussian PSFs with standard deviations of $0.12$~arcsec and $0.7$~arcsec for the HST imaging and HSC imaging, respectively. The background noise is assumed to follow a Gaussian distribution with a standard deviation of $0.05$ for HST and $0.12$ for HSC images. We also add Poisson noise to the images assuming an exposure time of 2100 seconds.

\begin{figure}
    \centering
    \includegraphics[width=1.0 \linewidth]{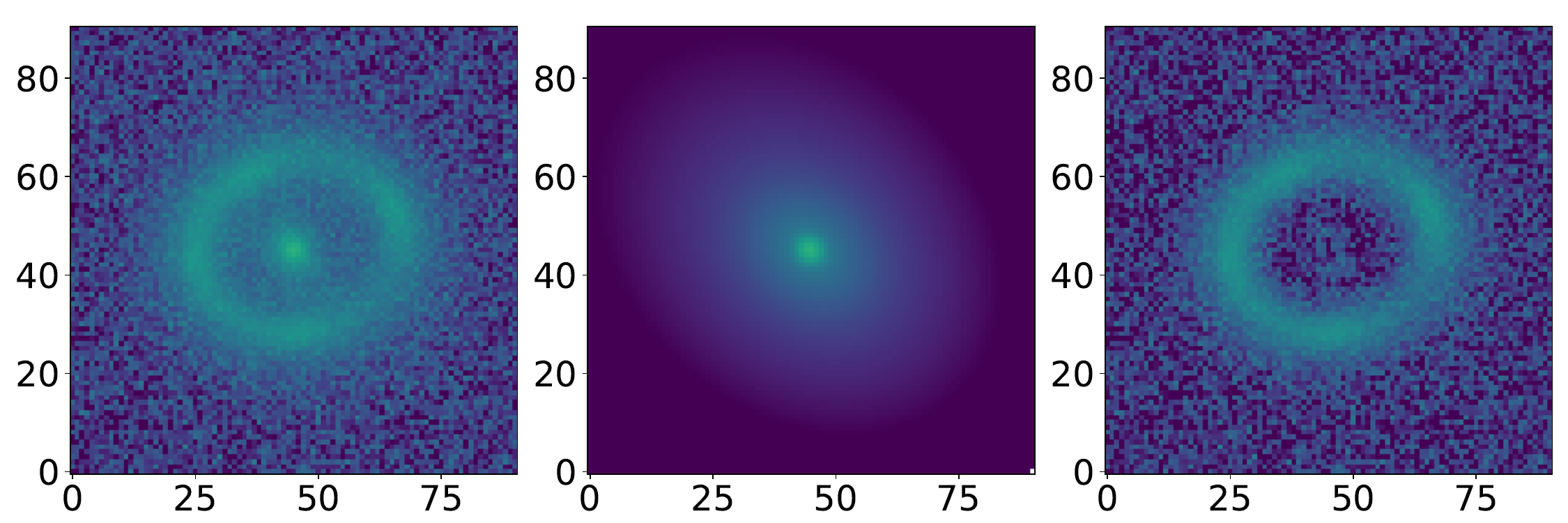}
    \includegraphics[width=1.0 \linewidth]{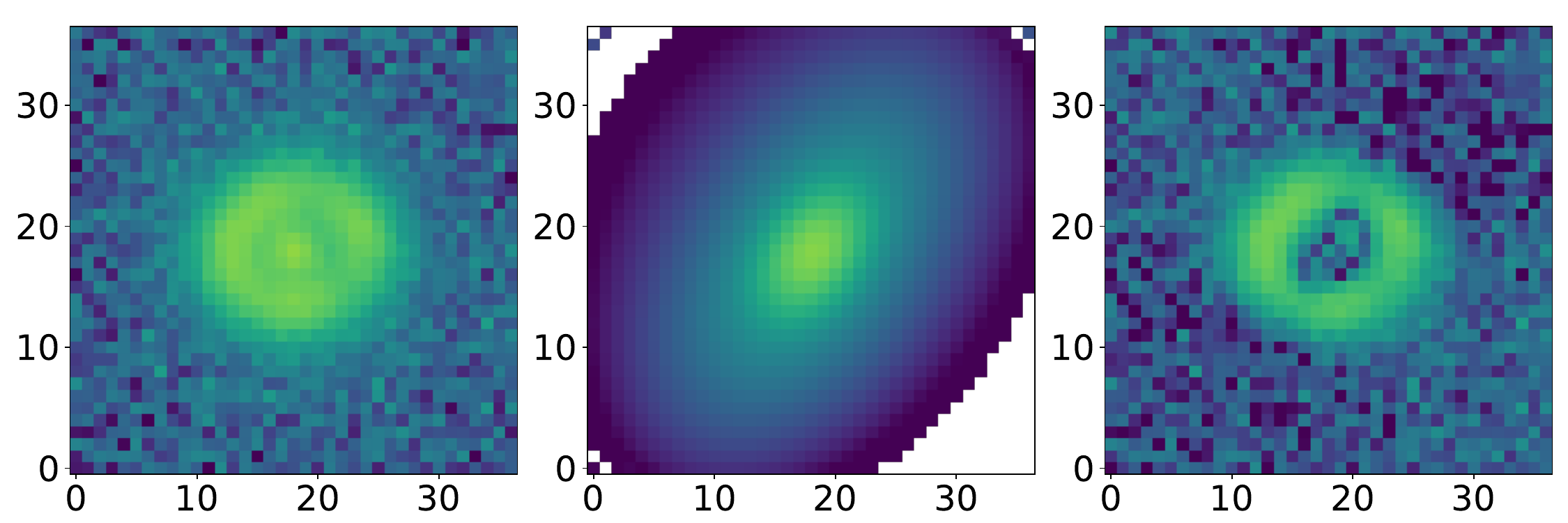}
    \caption{Lens light subtraction of a simulated HST ({\it top}) and HSC ({\it bottom}) lens system by \texttt{imcascade}.  {\it Left:} Simulated lens system with lens galaxy at the center. {\it Middle:} Best fit model for lens galaxy. {\it Right:} Residue image after subtracting the lens galaxy in the center. }
    \label{fig:HST_lens_light_subs}
\end{figure}

\subsection{Lens light subtraction}
We use \texttt{imcascade} \citep{Miller_2021} for subtracting the light profile of the lens galaxy. It adopts the Multi-Gaussian Expansion (MGE) method to represent the galaxy's light profile. Specifically, the PSF and light profile are both represented by a linear combination of Gaussians. Here, each Gaussian is defined by the standard deviation, and weight factor, which represents the fraction of contribution of a Gaussian to the sum of Gaussians. This is referred to as the Gaussian Mixture Model (GMM). The projected light distribution of the galaxy is modelled with $N$ components and written as 

\begin{equation}
    G(x, y) = \sum_{i=0}^{N} \frac{a_i}{2 \pi q \sigma_{i}^2} 
\exp \left[ 
-\frac{x'^2 + \left( \frac{y'}{q} \right)^2}{2 \sigma_{i}^2}
\right]
\end{equation}
with
\begin{equation*}
    x' = (x-x_{0})\cos(\phi) + (y-y_{0})\sin(\phi)
\end{equation*}
\begin{equation*}
    y' = -(x-x_{0})\sin(\phi) + (y-y_{0})\cos(\phi)\,.
\end{equation*}
Here, it is assumed that all Gausssians have the same central positions $(x_{0},y_{0})$, position angles $\phi$ and axis ratios $q$. The parameters $a_{i}$ and $\sigma_{i}$ are the weight and the standard deviation of $i$th Gaussian, respectively. Similarly, the PSF is also represented as a series of Gaussians. The observed light distribution will be the convolution of the PSF and $G(x,y)$. The standard $\chi^{2}$ minimization technique is used  to find the best-fit model to the data and obtain the parameters for a given number of Gaussian components. 

\begin{figure}[h]
    \centering
    \includegraphics[width=\linewidth]{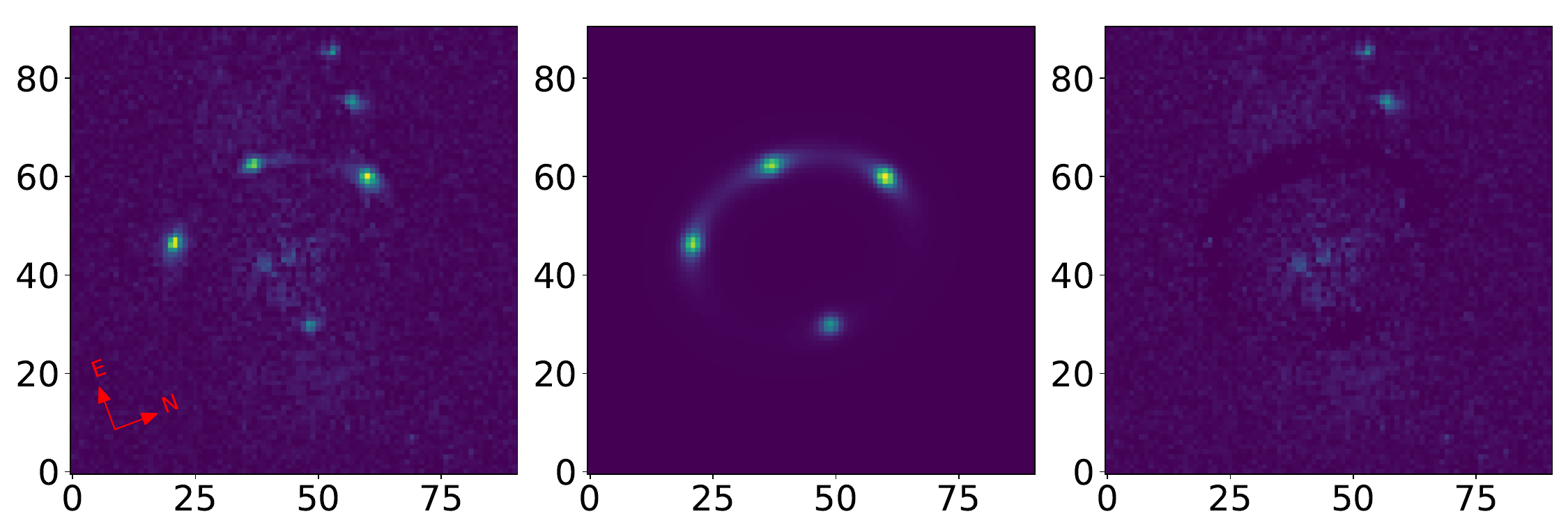}
    \includegraphics[width = \linewidth]{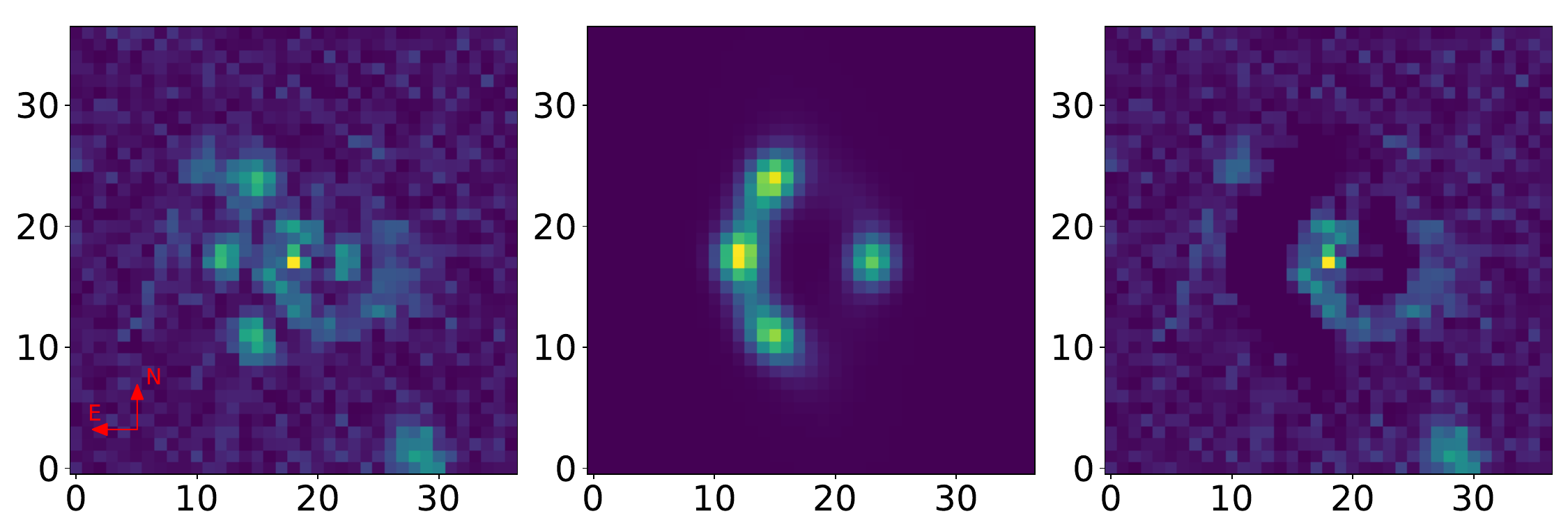}
    \caption{Modeling of \texttt{HSCJ230335+003703} for HST imaging in F814W (\textit{top}) and HSC imaging in the $i$~band. (\textit{bottom}). \textit{Left}:  Image showing the lensed features after lens light subtraction. \textit{Middle}: Best-fit model. \textit{Right}: Residual image after subtracting the lens model from the data.
    }
    \label{fig:Lens_model_from_PE}
\end{figure}

Some of the lensed images tend to be sufficiently close to the lens galaxy making the subtraction of lens light profile inefficient. In such cases, we mask the lensed images and then use \texttt{imcascade} to model the lens light. Once we obtain the best model of the light profile, we subtract it out to get the residual image. In Fig.~\ref{fig:HST_lens_light_subs}, we show an example of simulated lens system, matched to the resolution and image quality of HST (top) and HSC (bottom), for which the lens light is subtracted along with the residuals.

\begin{figure*}
    \includegraphics[scale=0.22]{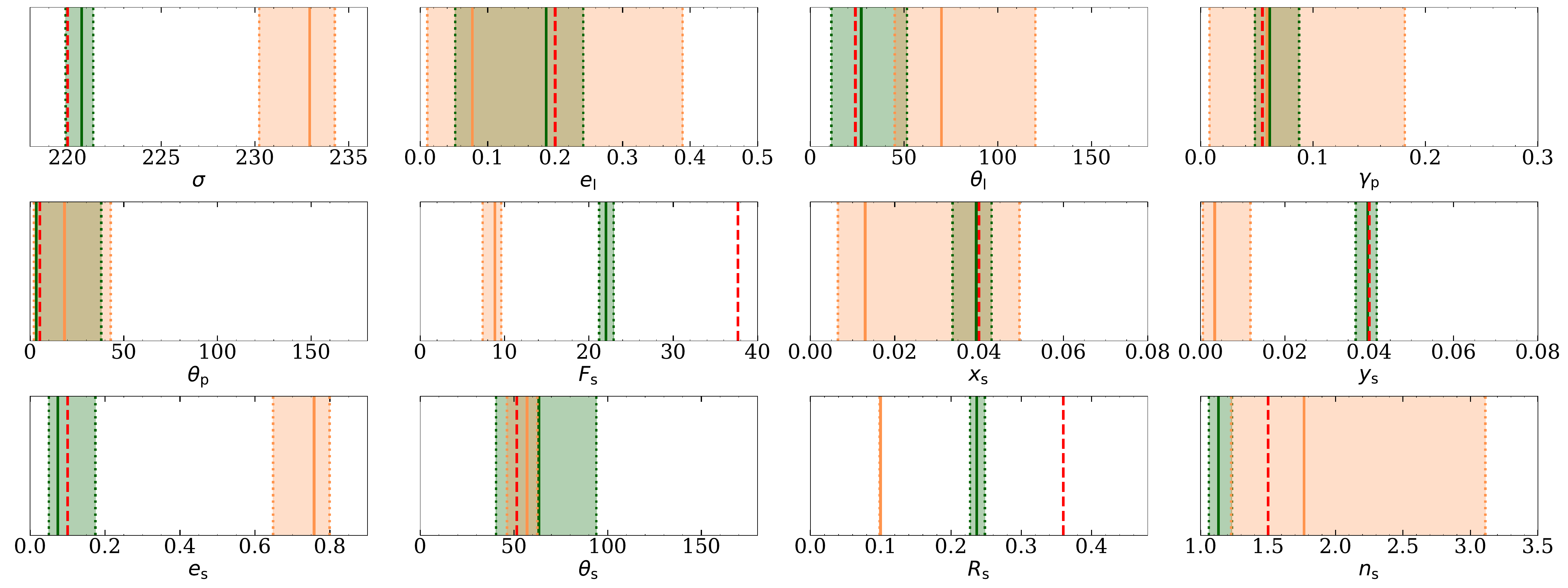}
    \caption{Inferred lens and source parameters from model fitting using DE for a simulated lens system in mock HST and HSC imaging.
    \textit{Green} color show the inferred parameters with 3~$\sigma$ confidence interval for mock HST.
    \textit{Orange} color shows the inferred parameters with 3~$\sigma$ confidence interval for mock HSC. \textit{Dashed red} line corresponds to the injected value of the corresponding parameters.}
    \label{fig:HST_PE}
\end{figure*}

\subsection{Lens mass modeling}
We use the lens-light subtracted residual images to find the best-fit lens mass model. Here, we use the SIE with an external shear to describe the mass distributions of the lens galaxies. To find the best-fit parameters of the mass model, we use an optimization algorithm called DE \citep{journals_jgo_StornP97,Qiang2014AUD}. DE is a population-based optimization algorithm used to solve complex multi-dimensional problems. It is a type of evolutionary algorithm, which iteratively improves a set of candidate solutions with respect to a given objective function. For more details see Appendix \ref{app:diff_evol}. The lens mass model and background source model together has 12 parameters as described below. The lens mass model parameters include the velocity dispersion
($\sigma_{\rm l}$), ellipticity ($e_{\rm l}$) and position angle ($\theta_{\rm l}$) for the lens whereas a strength ($\gamma_{\rm p}$) and the position angle ($\theta_{\rm p}$) for the external shear. The source galaxy parameters include the flux ($F_{\rm s}$), angular position ($x_{\rm s}, y_{\rm s}$), ellipticity ($e_{\rm s}$), position angle ($\theta_{\rm s}$), effective radius ($R_{s}$) and S\'ersic index ($n_{\rm s}$).
 
We experimented with the convergence of the DE algorithm by running it many times with different seeds. We find that the distribution of parameters doesn't change appreciably if we run the DE algorithm more than about 500 times. To obtain a robust estimation of the parameters, we choose to run the DE algorithm 600 times with varied initial seeds for each simulated lens.

In Fig.\ref{fig:Lens_model_from_PE}, we show a real lens system \texttt{HSCJ230335+003703} \citep{Wong_2018,refId0,10.1093/mnras/stae2442} and the best models obtained for the HST (top) and HSC (bottom) imaging using DE along with the residual images.

\begin{figure*}[htp]
\centering
    \includegraphics[scale = 0.21 ]{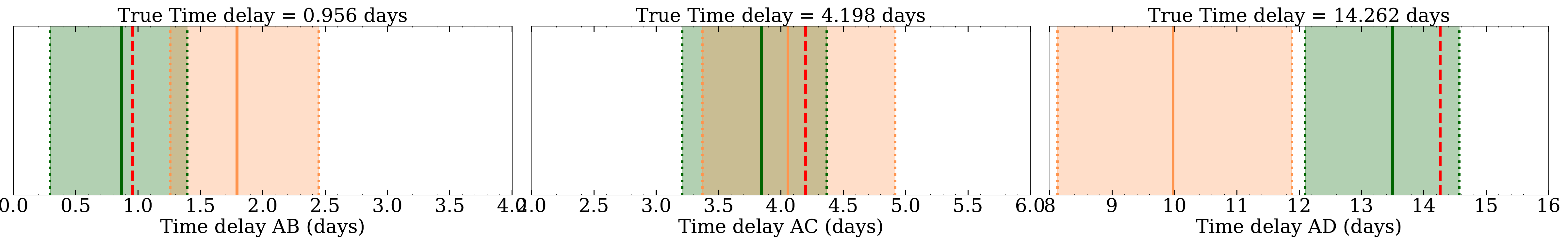}
    \caption{Time delays predicted for the same lens system observed in HST (green) and HSC (orange) with the 3~$\sigma$ confidence interval. The dashed line (red) corresponding to the true time delay.
    }
    \label{fig:Time_delay_plots}
\end{figure*}

\section{Results and Discussion} \label{results}

We present the results of lens modeling for a total of 16 mock lens systems analyzed for both HST and HSC data. Additionally, we study one real lens system selected to have both HST and HSC imaging.

For each lens system, we obtain the median and 3~$\sigma$ error bars from the 600 DE runs for our lens and source model parameters. We then use \texttt{glafic} to compute the corresponding median time delays and error bars. 

As an example, we show the inferred parameters for one of the mock lenses simulated to mimic  HST (green) and HSC (orange) imaging in Fig.\ref{fig:HST_PE}.

Firstly, we find that apart from the ${F_{\rm s}}$, ${R_{\rm s}}$ and ${n_{\rm s}}$, almost all parameters are well recovered for HST mocks. From the model fitting of different lens systems, we observed that these $3$ parameters are generally not well recovered. We believe that the reason for this is the degeneracies present among these parameters. 
Additionally, $\sigma_{\rm l}$ and ($ x_{\rm s}, y_{\rm s}$) are accurately estimated for HST as compared with the HSC. These parameters are especially important in calculating time delays. Therefore, time delays calculated for HST mocks are expected to be more accurate than for HSC mocks.

For the same lens system observed through HSC, we can clearly see that many inferred parameters are biased and are not well recovered. We believe that the principal reason for this is the low resolution of the telescope. Since $ \sigma_{\rm l}$ and ($x_{\rm s},y_{\rm s}$) are also not well recovered, the calculated time delays are also expected to be biased. 

\begin{figure*}[htp]
    \centering
    \includegraphics[scale = 0.50]{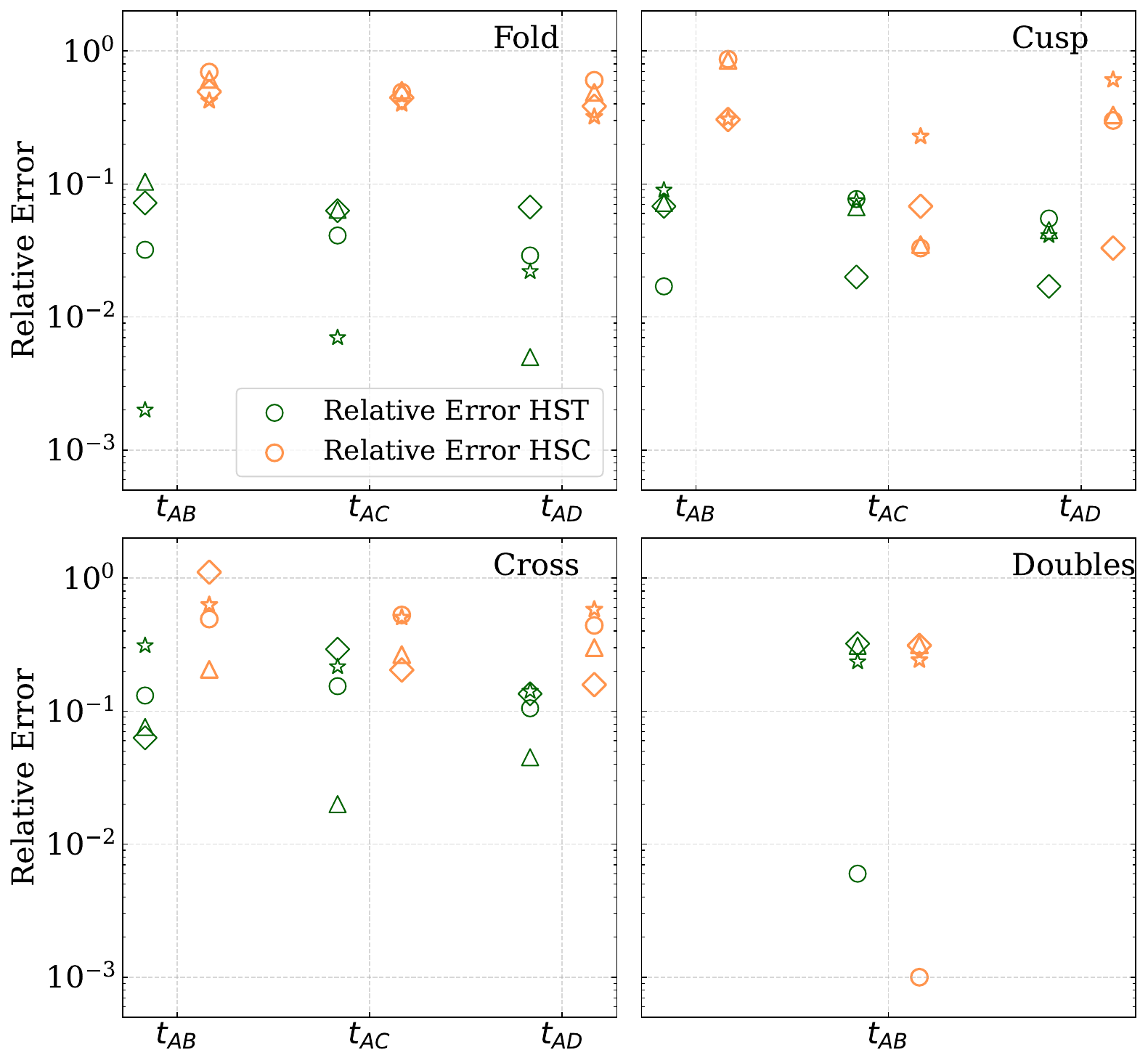}
    \caption{Relative error of time-delays for systems observed by HST (\textit{green}) and HSC (\textit{orange}). Different shapes corresponds to different lens systems.}
    \label{fig:rel_err}
\end{figure*}

\begin{figure*}[htp]
    \centering
    \includegraphics[scale = 0.50]{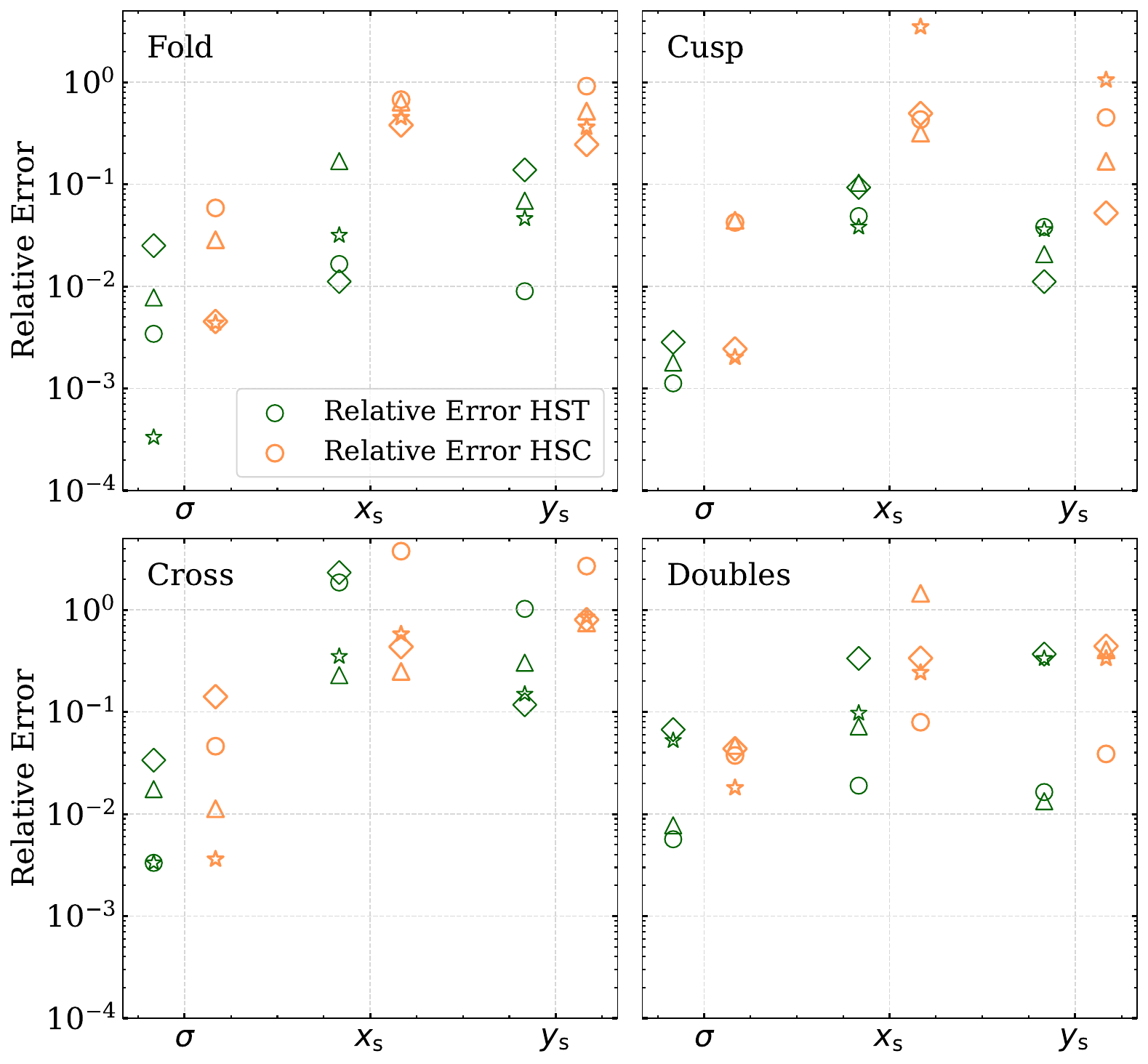}
    \caption{Relative error of $\sigma$, $x_{\rm s}$ and $y_{\rm s}$ for systems for mock HST (\textit{green}) and HSC (\textit{orange}) lenses. Different shapes corresponds to different lens systems.}
    \label{fig:rel_err_velocity}
\end{figure*}

In Fig~\ref{fig:Time_delay_plots}, we show the median of the time delays and 3~$\sigma$ error bars for image pairs AB, AC and AD for the same simulated lens system as in Fig.\ref{fig:HST_PE}. The lensed images are labelled alphabetically with increasing time delay such that image A is the first arriving image. The time delays in any pair of images are with respect to image A. Therefore, AB is the time delay between A and B. As expected from the accurately inferred $\sigma_{\rm l}$ and ($ x_{\rm s}, y_{\rm s}$) for the HST, the true time delay lies within $3\sigma$ of the predicted value. Meanwhile, for the same lens system, the time delays predicted from HSC mocks can be biased for some of the image pairs. 

We repeat the same analysis for the 16 mock lens systems generated for both HST and HSC. 
As described in section~\ref{ssec_gen_sim}, we have four distinct lenses for each double as well as for quads, in the fold, cusp and cross configuration.  The varied configurations and multiplicity of the lenses considered in our analysis will be useful in identifying the lenses which tend to produce less biased or more accurate time delays. 


In Fig.~\ref{fig:rel_err}, we plot the relative errors in time delays for all $16$ different lens systems sorted according to their configurations and also their relative errors in time delays for both HST and HSC mocks. A general trend that we observe is that for quads, the time delays calculated using HST are much more accurate than those using HSC.  We believe that the principal reason is the biased inference of $\sigma_{\rm l}$ and ($ x_{\rm s}, y_{\rm s}$) which are especially important in calculating time delays. This can be seen from Fig.~\ref{fig:rel_err_velocity} where if the relative errors in $\sigma_{\rm l}$ and ($ x_{\rm s}, y_{\rm s}$) are larger, then the corresponding relative error in the time delays (see Fig.~\ref{fig:rel_err}) also tends to be larger for the quads.

Out of all the configurations, the time delays calculated for a cross are much more biased than other configurations. In a cross configuration, the images are much closer to the lens, because of which the images are not well resolved and in some situations can be hidden by the lens light. Even though we are subtracting the lens light, because the images are closer to the lens, an improper lens light subtraction may affect the light from the images and thus the fitting to the lens model. We believe that this might make the prediction of the time delays for a cross challenging. Even within cross configurations, we find that relative errors in time delays are smaller from the HST mocks as compared to the HSC mocks. 

For the case of doubles, we find that the relative error in the time delays from HSC is comparable to that of HST. In the case of doubles, the images are further away from the lens and well separated, hence are not much affected by the lens light. It can be seen from Fig.\ref{fig:rel_err_velocity} that for the doubles in HST, the relative errors in $\sigma_{\rm l}$ and ($ x_{\rm s}, y_{\rm s}$) are of the same order of magnitude as HSC, so the corresponding relative error in the time delays is also comparable between the HST and HSC mocks in Fig.\ref{fig:rel_err}. Since we know the positions of the images more accurately, the relative error in time delays is comparable between HSC and HST. 

\begin{figure*}[ht]
\centering
    \includegraphics[scale = 0.21 ]{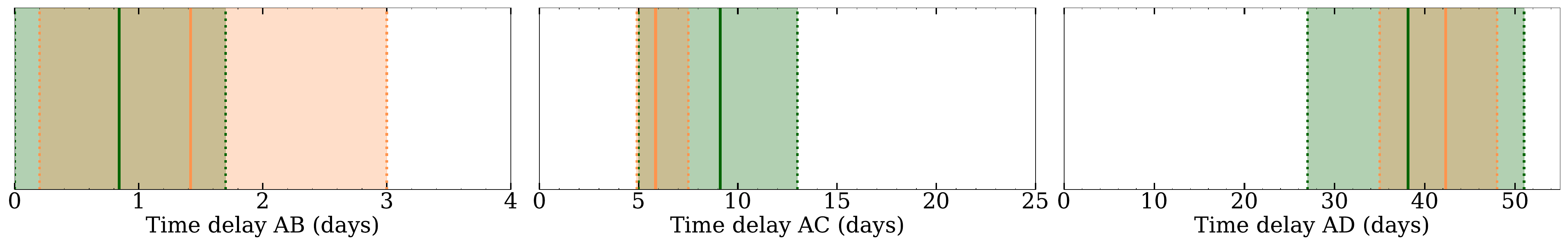}
    \caption{Time delay distribution inferred for a real lens system observed in HST and HSC.}
    \label{fig:Time_delay_plots_real_system}
\end{figure*}

Now, we describe the results of our analysis on \texttt{HSCJ230335+003703}, a real lens system (see Fig. \ref{fig:Lens_model_from_PE})
which is observed by both HST and HSC. This allows us to compare the predicted time delays from the two data. The Fig.~\ref{fig:Lens_model_from_PE} shows the HST and HSC images of the lens system after lens light subtraction, the best-fit model, and the residual image. The Fig.\ref{fig:Time_delay_plots_real_system} shows the predicted time delays and the 3~$\sigma$ error bars for this system from HST and HSC images. We can see that the maximum difference in time delays calculated from HSC and HST is $\sim 5 $ days. This difference between the predictions of HST and HSC images can be largely attributed to the poor resolution of HSC over HST, but other factors like imperfect lens light subtraction and nature of the configuration cannot be ignored. 

\section{Summary and Conclusion}\label{conclusion_section}

Following our proposal of a novel GW early-warning method that exploits gravitational lensing of BNS/NSBH mergers that produce GWs with EM counterparts \citep{Magare_2023}, we assess the feasibility of our method to make early warning predictions for subsequent GW (including EM) counterparts. The key to the success of our proposal involves predicting time delays from the lensed images of the galaxy that hosts the BNS/NSBH using optical data. These predictions need to be rapid and with sufficient accuracy to be able to find the subsequent counterparts.

As most of the immediate follow-up optical observations tend to be with ground-based facilities, we investigate the impact of the low resolution and atmospheric seeing on the time delay predictions. We therefore simulate galaxy-scale lens systems at both high and low angular resolutions in a single broad band filter ($i-$band). The data are matched to the imaging quality and resolution of ACS camera on the HST and HSC camera on the Subaru telescope. 

We study doubles and quads as these image multiplicities are the most standard for galaxy-scale lenses. Since quads show three typical image configurations with distinct qualitative features, studying them can give insights into any systematic biases and uncertainties. Thus, we repeat the analysis for the standard quad configurations, namely, fold, cusp and cross. In total, we simulate 16 unique lenses, four for each of the quads and doubles.

We perform lens light subtraction to minimize contamination from the lens light during the modeling of the lensed images. We fit an SIE mass model to the lens-light-subtracted images and predict the time delays from the best-fit model. The differential evolution (DE) optimisation routine is run 600 times for each mock lens which is then used to determine the uncertainties. The HST and HSC mocks for a given lens system are modelled independently. 

Comparing the inferred lens and source parameters for HST vs HSC mock lenses, we find that almost all parameters are well recovered for the HST mocks within 3~$\sigma$ error bar. In contrast, the inferred parameters tend to be biased for HSC mocks. As a result, the predicted time delays are accurate for HST, but can be biased for ground-based imaging such as HSC.

Next, for doubles, the relative errors in the time delays from HSC mocks are comparable to those of HST mocks, with no bias noted in either. Conversely, for quads, the relative errors from HST data are much smaller than that from HSC. Out of the three quad configurations, cusp have smaller relative errors when using HSC images and can be a better candidate for getting accurate time delays when using ground-based telescopes. Furthermore, if we have prior information about the velocity dispersion of the lens, this can further improve the time delay predictions.

We chose DE as a faster alternative to the conventional Markov Chain Monte Carlo (MCMC) approach. We find that the wall-clock time on a single CPU core for lens mass modeling with 12 free parameters is $\sim6-10$~hrs for a single DE run, while for MCMC it is around ~$\mathcal{O}(10)$ hrs for convergence. Since we want to predict the time delays of the images as fast as possible, we find DE-based optimisation is better compared to the conventional MCMC approach without significantly compromising the accuracy\footnote{This is a general remark based on our experience with MCMC analyses rather than a direct comparison which was not performed in this work.}.

Our work demonstrates the feasibility of our early-warning approach using realistic simulated galaxy-scale lenses. We are able to predict time delays in a few hours if optical imaging of the lensed host galaxy is available. Our analysis is applicable to the upcoming Rubin observatory's Legacy Survey of Space and Time which will be finding $\mathcal{O}(10000)$ lenses and will also probably be used for following up the candidate BNS/NSBH triggered by the LVK's low latency alert software infrastructure.


\section*{Acknowledgements}
We thank Arshi Ali for help with providing scripts and inputs related to differential evolution optimisation and Saee Dhawalikar for helpful discussions. S.J.K. gratefully acknowledges support from ANRF/SERB grants SRG/2023/000419 and MTR/2023/000086. We acknowledge the use of IUCAA LDG cluster Sarathi for the computational/numerical work. This material is based upon work supported by NSF’s LIGO Laboratory which is a major facility fully funded by the National Science Foundation. 
\vspace{5mm}

\textit{Software}: \texttt{NumPy} \citep{vanderWalt:2011bqk}, \texttt{SciPy} \citep{Virtanen:2019joe}, \texttt{astropy} \citep{2013A&A...558A..33A, 2018AJ....156..123A}, \texttt{Matplotlib} \citep{Hunter:2007}, \texttt{jupyter} \citep{jupyter}.

\bibliography{references}

\appendix
\section{Differential evolution algorithm}
\label{app:diff_evol}

\begin{figure*}[htp]
    \centering
    \includegraphics[scale=0.27]{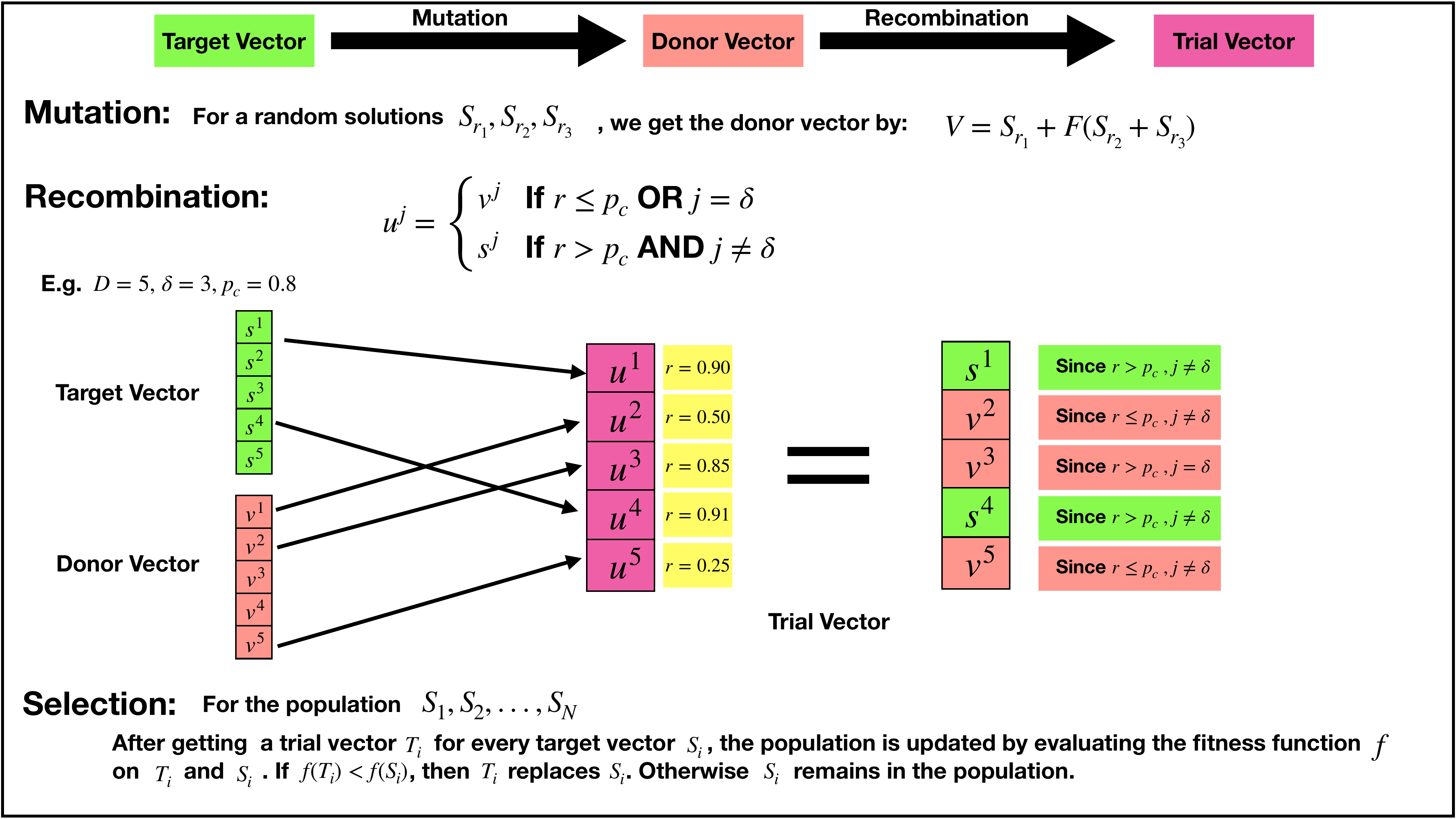}
    \caption{ A flowchart of Differential Evolution algorithm. Given a population size of N, the algorithm starts with initial solutions $S_{1}, S_{2},...,S_{N}$ where each solution is D-dimensional vector, called Target vector. Each solution then undergoes mutation followed by recombination. In mutation, 3 random non-identical target vectors are picked and combined to get a Donor vector ($V$) as shown above. Here F is a scale factor between 0 and 2. In the recombination step, the Target vector and the Donor vector are recombined using the above method to get the Trial vector (T). Above figure illustrates an example of recombination for $D=5, \delta = 3$, which is a random integer between (1 and D) and $p_{c} = 0.8$, which is called the crossover probability. Here $s^{i}, v^{i}, u^{i}$ are the $i$th component of Target, Donor and Trial vectors respectively. Once the Trial vectors are generated corresponding to each Target vector, a greedy selection rule is employed in which the population is initial solution ($S_{1}, S_{2},...,S_{N}$) is updated as illustrated above.}
    \label{fig:diff_evol}
\end{figure*}

DE algorithm starts with a given population size $N$, which contains the solutions $S_{1}, S_{2},..S_{N}$, each solution is a $\rm D$-dimensional vector called the Target vector. In our case, we have $D=12$ and corresponds to parameters of the SIE $+$ external shear model.

Each solution then undergoes Mutation followed by Recombination. In the mutation step, 3 random non-identical Target vectors $(S_{r1},S_{r2},S_{r3})$ are combined to get a donor vector $V = S_{r1} + F(S_{r2}+S_{r3})$, where $F$ is a scale factor between $0$ and $2$. In the recombination step, the Target vector and the Donor vectors are combined to get a trial vector ($T$). Each element of the Trial vector is either an element of the Target vector or of the Donor vector, this is decided by the algorithm illustrated in Fig.$\ref{fig:diff_evol}$. The fraction of elements from the Donor vector to the Target vector depends on crossover probability ($p_c$). $r$ is the random number between $0$ and $1$, and is evaluated $\rm D$ number of times i.e. for each element of Trial vector. $\delta$ is the random integer between $1$ and $\rm D$, and is evaluated for every Target vector. Fig.$\ref{fig:diff_evol}$ shows an example of recombination step with $\rm D=5$, $\delta = 3$ and $p_c = 0.8$.

Once we generate the Trial vector for each Target vector, a greedy selection rule is applied. For each Trial vector $T_i$ and Target vector $S_i$, the objective function $f$ is evaluated and if $f(T_i) < f(S_i)$, then $T_i$ replaces $S_i$ in the population of candidate solutions, otherwise $S_i$ remains in the population. The above steps, from Mutation to Selection, are repeated again and again for the solution to converge at the minimum of $f$.

\end{document}